\documentstyle[11pt,newpasp,twoside,epsf]{article}
\def\edcomment#1{\iffalse\marginpar{\raggedright\sl#1\/}\else\relax\fi}
\reversemarginpar

\begin{document}
\title{The Morphology of Extremely Red Objects}
 \author{Massimo Stiavelli}
\affil{Space Telescope Science Institute, 3700 San Martin Dr., Baltimore MD 21218, USA}
\author{Tommaso Treu}
\affil{Scuola Normale Superiore, Piazza dei Cavalieri 7, 56126 Pisa, Italy}

\begin{abstract}
We briefly discuss our survey of Extremely Red Objects (EROs) with
HST/NICMOS and deep ground-based images and summarize its results. We
use the high-resolution NICMOS images to sub-divide EROs into
morphological classes, tentatively arguing that about 60 \% of EROs
(18 out of 30) are early-type galaxies. About 20 \% (6 out of 30
objects) appear to be flattened disk systems. Another 10 \% (3 out of
30) have an irregular morphology. Thus, combining disk and irregular
systems, our data are compatible with 30 \% of EROs being dusty
starbursts. A small fraction (3 out of 30) appear point-like even on
the HST images and could therefore be either compact high redshift
galaxies or stars in the Milky Way halo.
\end{abstract}

\section{Introduction}

The advent of high quality near-infrared detectors has allowed a
veritable breakthrough in the study of extremely red objects
(hereafter EROs), {\it i.e.} galaxies characterized by $R-K$ colors of
5 or redder. Clearly, a natural interpretation of their red colors is
that EROs are elliptical galaxies with relatively old stellar
populations at a redshift between 1 and 2.  This has been confirmed by
Spinrad et al. (1997) for the ERO with regular morphology LBDS 53W091.
However, Graham \& Dey (1996) have shown that another ERO, HR10,
actually has a spectrum with emission lines. Cimatti et al. (1998)
definitely confirmed the dusty nature of HR10. Clearly, on the basis
of colors alone it is not possible to discriminate between a dusty
starburst or an old stellar population at $z>1$. On the other hand,
spectroscopy of a large sample of EROs is, unfortunately, prohibitive
even with 8-meter class telescope ({\it e.g.} Cimatti et al. 2000).

One important difference between LBDS 53W091 and HR 10 is their
morphology.  At HST resolution, the former appears regular while the
latter has an irregular morphology.  This has led us to adopt as a
working hypothesis the assumption that morphology with HST can help
identify the nature of EROs. Therefore, we have carried out a survey
of EROs in fields where HST/NICMOS images were available.  In Section
2, we describe our survey and some basic results. Section 3 is devoted
to a discussion of morphologies. Section 4 sums up.

\section{An ERO survey with HST/NICMOS}

We have carried out a survey of EROs over an area of 13.74
arcmin$^{2}$ using HST/NICMOS F160W images, for a total of 23
fields. We have complemented the near-IR data with either WFPC2 F606W
images or ground-based R-band images (see Treu \& Stiavelli 1999,
hereafter TS99, for more details). We built a number of Bruzual \&
Charlot (1993, and 1996 unpublished) models to assess the range of
colors displayed by a variety of star formation histories. We focussed
on two star formation histories that may be relevant for high-redshift
elliptical galaxies: {\it i)} a stellar population 1 Gyr old at any
given redshift, and {\it ii)} a stellar population formed as a single
burst at z=10. EROs were selected according to two color criteria
determined on the basis of these models. The first, broader, criterion
is aimed at selecting elliptical galaxy candidates at $z\geq1$ and
defines our basic ERO sample. Its color criteria are R-H$_{160} > 3.2$
and V$_{606}$-H$_{160}>3.8$ (here and in the following we will use
magnitudes in the AB system.) The second criterion selects candidate
elliptical galaxies at $z\geq1.5$ and corresponds to R-H$_{160} > 4.0$
and V$_{606}$-H$_{160}>4.5$.

\begin{figure} 
\plotfiddle{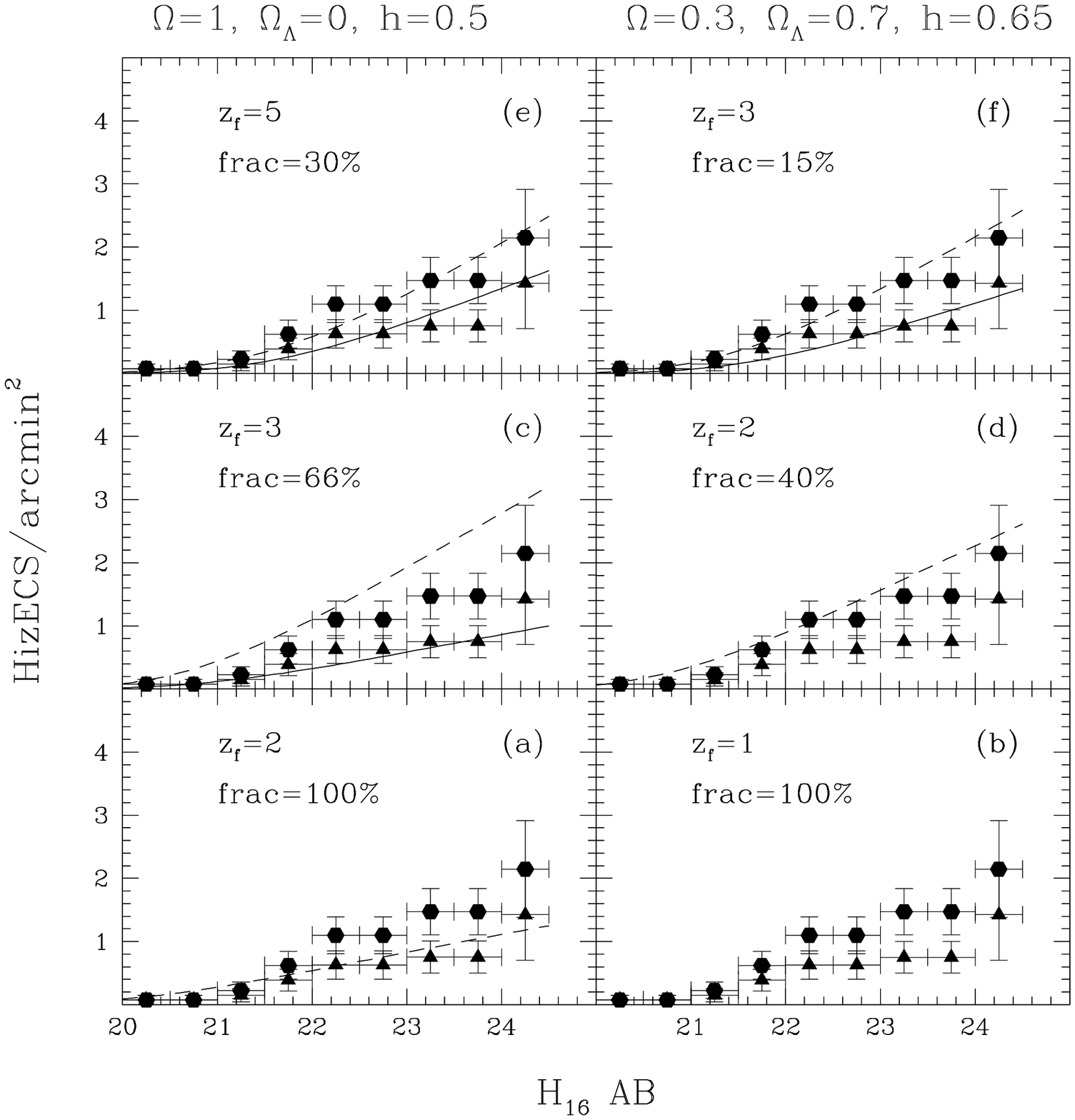}{7.0truecm}{0.0}{50.0}{35.0}{-150.0}{-50.0}
\caption{We plot the observed luminosity function of candidate
high-redshift elliptical galaxies (HiZECs) at $z\geq1$ (hexagons) and
$z\geq1.5$ (triangles) as a function of observed magnitude in the
F160W filter. The data are compared to the luminosity function
obtained for local early-type galaxies formed as a single burst of
star formation (solid line for $z\geq1.5$, dashed line for
$z\geq1.0$). The left panels and the right panels refer to different
cosmologies. Each panel lists the formation redshift (z$_f$) and the
fraction of early-type galaxies required to fit the data.}
\end{figure}

We have identified a total of 30 EROs.  The completeness of our data
set was verified by carrying out a large set of Montecarlo simulations
showing that at the faintest level we are still more than 90 per cent
complete. The luminosity function that we derive extends down to
H$_{160} = 24.5$ (only for a fraction of the field, see TS99 for a
detailed list of the area covered at each depth.) At this depth we
detect $\sim 3\pm1$ EROs arcmin$^{-2}$.

Each ERO was morphological classified by visual inspection
independently by the two authors. In general there was a good
agreement between the two independent classifications by the authors
but in four cases there was a discrepancy. These cases were looked
in more detail until a common classification could be agreed upon. The
final breakdown of morphologies are given in Table 1. Those objects
classified as early-type galaxies (E/S0) were then used to derive a
luminosity function to be compared with the one observed at low
redshift under the assumption of passive evolution. The number density
of {\it regular} EROs in our survey at the limit of H$_{160} = 24.5$
is $\simeq 2\pm0.5$ arcmin$^{-2}$.  The derived luminosity function is
compared to the observed one in Figure 1. The number of red galaxies
that we find requires the early formation of at least some early-type
galaxies. However, our data do not support the formation at
high-redshift of all early-type galaxies, {\it i.e.} we see a deficit
of red galaxies. Each panel in the figure indicates the fraction of
early-type that could have formed at the given redshift. This fraction
ranges from 15 to 66 per cent depending on cosmology and formation
redshift (TS99). This discrepancy could be interpreted in a number of
ways:

\noindent {\it i)} all ellipticals are already in place at high
redshift but we are only seeing those that have not been polluted by
younger populations. In fact, only a very modest fraction of young
stars would be sufficient to make an elliptical appear bluer than our
selection criteria;

\noindent {\it ii)} only some of the early-type galaxies were in place at
$z\geq2$. As an example, lenticular galaxies could have formed later
while giant ellipticals were already present at $z=2$;

\noindent {\it iii)} our error bars are underestimated due to the
clustering of EROs seen, {\it e.g.} by Daddi et al. (2000), and our
results are thus compatible with all early-type galaxies forming at
high redshift.

Clearly it is difficult at this stage to determine which of these
alternatives (or their combinations) is correct. However, our data
strongly suggest that at least some elliptical galaxies were already
in place at relatively high redshift.

\section{Morphologies}

In the course of our morphological analysis we have binned EROs in
four classes: {\it i)} E/S0, characterized by regular morphology, {\it
ii)} disk, characterized by high apparent flattening, {\it iii)}
irregular, and {\it iv)} point-like. The point-like objects could be
red stars in the Milky Way halo, small elliptical galaxies, or compact
regions embedded in low surface brightness objects. The number of
point-like objects that we detect (3) is well within the limits
expected for brown dwarfs in the Galactic halo ({\it e.g.}  Flynn et
al. 1996).

In the analysis of the luminosity function of early-type galaxies we
have included only those objects belonging to the E/S0
category. Clearly there could be a contamination introduced by face-on
disk EROs. However, the E/S0 EROs tend to be characterized by a
R$^{1/4}$ light profile ({\it e.g.} Stiavelli et al. 1999); this is in
contrast to the low redshift luminous disk galaxies none of which has
an R$^{1/4}$ light profile. This argues in favor of little or no
contamination of the E/S0 sample from disk systems.  This also implies
that face-on disk systems would have not been classified as E/S0 EROs.
In general, these face-on disks would likely be observed as irregular
EROs or even drop out from the EROs sample if, {\it e.g.}, the red
colors of edge-on disk EROS are due to dust absorption through the
line of sight of the disk.

\begin{table}
\begin{center}
\caption{Morphological classification of EROs in the TS99 sample}
\begin{tabular}{lc}
\tableline
Class & Number \\
\tableline
E/S0  & 18\\
disks & 6 \\
irregular & 3 \\
point like & 3 \\
\tableline
\tableline
\end{tabular}
\end{center}
\end{table}

\section{Discussion and Conclusions}

Our results suggest that some early-type galaxies were in place at
high redshift. We detect a deficit of red objects which may imply
either that some early-type galaxies formed later, or that
observational effects or interactions interve to make the colors bluer
and thus cause some early-type galaxies to drop from the E/S0 ERO
sample.

We find a number density of about 0.5 arcmin$^{-2}$ (at H$_{160} =
24.5$) of EROs that appear to be as flattened as a disk. They could be
massive disks already in place at $z>1$. We expect these objects as
well as the irregular EROs to be characterized by a dusty starburst
spectrum. 

The major limitation of our study is the relatively small area of the
survey especially in light of the large field-to-field fluctuations
discovered by Daddi et al. (2000) for the bright-end of the ERO
distribution. For this reason, we are planning to increase the area of
our survey by a factor four in the near future.

\end{document}